\documentclass{article}
\usepackage{graphicx}
\usepackage[a4paper]{geometry}
\usepackage{amssymb}
\usepackage{amsmath}
\usepackage{upgreek}
\usepackage[version=3]{mhchem}
\usepackage{placeins}

\begin{document}
\title{Forward proton detectors in heavy ion physics}
\author{
Rafa{\l} Staszewski,
Krzysztof Cie\'sla,
Janusz J. Chwastowski\\[7mm]
Institute of Nuclear Physics Polish Academy of Sciences\\
ul. Radzikowskiego 152, 31-342 Krak\'ow, Poland\\[7mm]
}

\maketitle
\begin{abstract}
  The forward proton detectors, already existing at the LHC, are considered in the context of heavy ion collisions.
  It is shown that such detectors have the potential to measure nuclear debris originating from spectator nucleons.
  The geometric acceptance for different nuclei is studied, and how it is affected by the motion of the nucleons in the nucleus and by the experimental conditions.
  A possibility of reconstructing the impact parameter of the collision from the measurement of the nuclear fragments is discussed.

\end{abstract}

\section{Introduction}

The Large Hadron Collider \cite{bib:lhc} is equipped with dedicated detectors allowing measurements of protons scattered in diffractive or electromagnetic interactions.
Since the scattering angles of such protons are very small, these detectors are installed very far away from the interaction point.
In addition, due to the use of the roman pot technology, they can be placed very close to the proton beams.

The LHC physics program is not solely devoted to the studies of the proton--proton interactions.
Possibilities of accelerating heavy ion beams have already resulted in many measurements of proton--lead and lead--lead collisions (see for example \cite{Proceedings:2019drx} and references therein).

\begin{figure}[thb]
  \centering
  \includegraphics{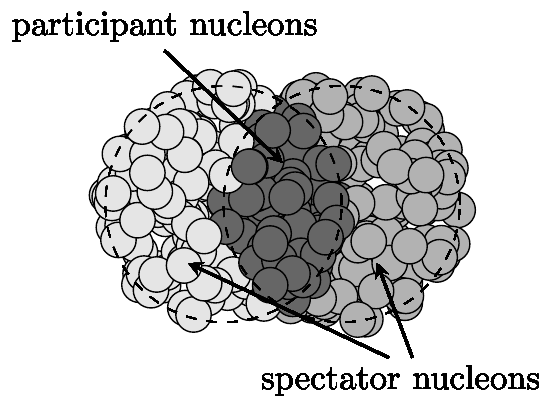}
  \caption{A schematic diagram of a ultra-relativistic heavy ion collision (view perpendicular to the relative velocities).}
  \label{fig:AA_interaction}
\end{figure}

An ultra-relativistic interaction of two heavy nuclei is sketched in Fig.~\ref{fig:AA_interaction}.
Typically, the impact parameter of the collision has a non-zero value and only a part of the nucleons constituting one nucleus collides with a part of the nucleons belonging to the other nucleus.
The nucleons actively participating in the interaction are called the participant nucleons (or participants),
in contrast to the spectator nucleons (spectators).

Since the time-scale of the ultra-relativistic ion collision is much shorter than that of the interactions within the nuclei,
the spectators are essentially left intact during the nucleus--nucleus collision.
They are scattered into the accelerator beam pipe and escape the acceptance of the central detectors, very much like diffractive protons.
The present work tries to answer whether and to what extent the forward proton detectors at the LHC could be used with heavy ion beams.

\FloatBarrier
\section{Forward proton detectors}
\label{sec:detectors}

\begin{figure}[t]
  \centering
  \includegraphics[width=\linewidth]{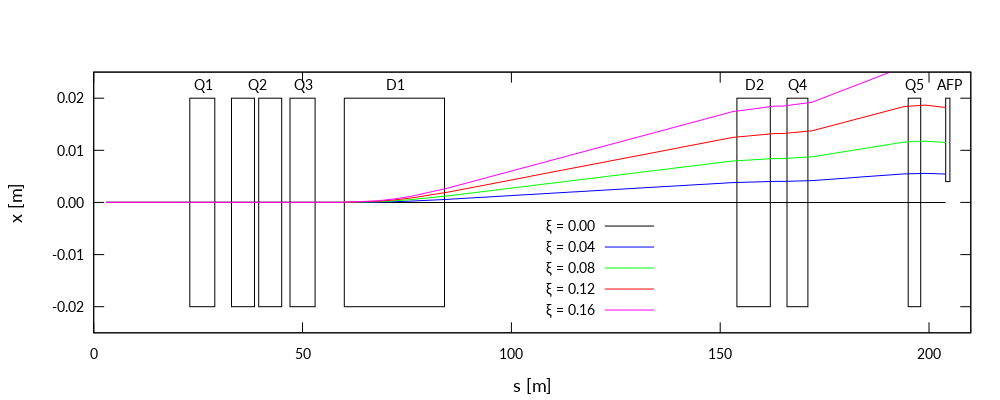}
  \caption{ 
  Trajectories of diffractively scattered proton in the LHC magnetic fields. 
  $x$ is the horizontal coordinate of the trajectory with respect to the nominal orbit,
  $s$ is the distance from the interaction point along the orbit,
  $\xi$ is the relative energy loss of the protons.
  }
  \label{fig:trajectories}
\end{figure}

Several systems of forward proton detectors are installed at the LHC: TOTEM~\cite{bib:totem}, CT-PPS~\cite{bib:pps}, ALFA~\cite{bib:alfa} 
and AFP~\cite{bib:afp}.
All these detectors are placed about 200 m away from their corresponding interaction points.
The ALFA detectors approach the beams in the vertical direction, the AFP and CT-PPS approach horizontally, while TOTEM has both types of detectors.
The present analysis takes the AFP detectors as an example for the simulations.
The results for other horizontal ones can be expected qualitatively similar.
The potential application of the vertical detectors is not to be covered within this work.

The AFP (ATLAS Forward Proton) detectors are a sub-system of the ATLAS experiment set-up located at the LHC Interaction Point 1. 
It consists of four stations -- two on each outgoing beam.
The near stations are placed 205 m and the far ones 217~m away from the interaction point.

Each AFP station consists of a roman pot mechanism allowing the horizontal insertion of the detectors into the accelerator beam pipe.
Each station contains a tracking detector of four planes of the 3D Silicon sensors \cite{bib:Si}. 
Additionally, the far stations can be equipped with quartz-based Cherenkov time-of-flight counters.
However, these counters are not relevant for the present study.

A diffractively scattered proton, before being detected in AFP, passes through the magnetic fields of seven LHC magnets.
The Q1 -- Q3 triplet of quadrupole magnets is responsible for the final focusing and the emittance matching of the beams, providing thus the high luminosity of collisions.
The two consecutive magnets are the dipoles: D1 and D2.
D1 separates the incoming and the outgoing beams, while D2 accommodates them within the corresponding beam pipes of the machine.
The last two quadrupole magnets, Q5 and Q6, are used to match the beam optics in the interaction region to the rest of the ring.

The momentum of the diffractively scattered proton slightly differs from that of the beam particles. 
In the interaction, the proton is scattered at a small angle and looses some part of its energy.
Scattering at small angles means that the distribution of the transverse momentum of the diffractively scattered protons is very steep.
Therefore, it is the scattered proton energy which mainly determines its trajectory and hence the position in the forward detector.
The transverse momentum of a typical magnitude only leads to a moderate smearing of this position. 

The lower energy of the diffractive proton means that the curvature of its trajectory in the magnetic field will be greater, which will cause the scattered proton to recede from the beam orbit. 
This property allows the measurement of such protons with detectors placed close to the beam. 
As an example, Fig.~\ref{fig:trajectories} shows trajectories of protons with different energy values, here specified by the relative energy loss $\xi = 1-E_{proton}/E_{beam}$.
Also, the LHC magnets and the AFP detector are depicted in this figure.

The kinematic range in which the measurements are possible can be quantified by the value of the geometric acceptance given as a function of the proton 
energy and its transverse momentum (averaged over the azimuthal angle). 
Naturally, the acceptance depends on the exact settings of the magnetic fields, the so-called machine optics, and on the position of the detector with respect to the beam.

\begin{figure}[htbp]
  \centering
  \includegraphics[width=0.5\linewidth]{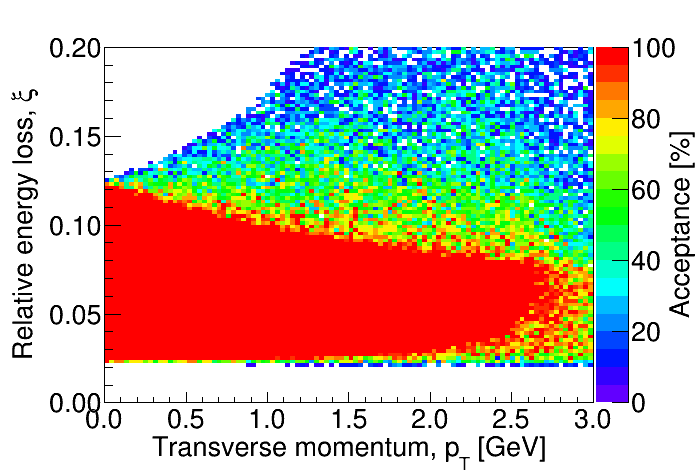}
  \caption{
  Geometric acceptance of the AFP detectors for diffractive protons.
  }
  \label{fig:proton_acceptance}
\end{figure}

Fig. \ref{fig:proton_acceptance} presents an example acceptance plot, calculated for the design LHC optics 
\cite{bib:lhcoptics}. The presented results were obtained using the Mad-X programme \cite{bib:madx}.
One can see that the AFP detectors can measure protons that lost between 2\% and 12\% of its energy and gained less than 
2.5~GeV of transverse momentum. 

\FloatBarrier
\section{Acceptance for spectator fragments}
\label{sec:acceptance}

The spectator nucleons after an interaction of ultra-relativistic heavy ions are left in a very peculiar state.
Before the collision the nucleons were a part of the nucleus, interacting with other nucleons belonging to it.
Then, the participating nucleons are ``taken away'' and the spectators are left as members of an unstable ensemble,
which subsequently decays into lighter fragments.
In order to assess the geometric acceptance of the detectors for the spectator fragments, it is assumed that all the nuclei lighter than the projectiles can possibly be produced without paying any attention to the abundances of particular isotopes.

\begin{figure}[htbp]
  \centering
  \includegraphics[page=2, width=0.5\linewidth]{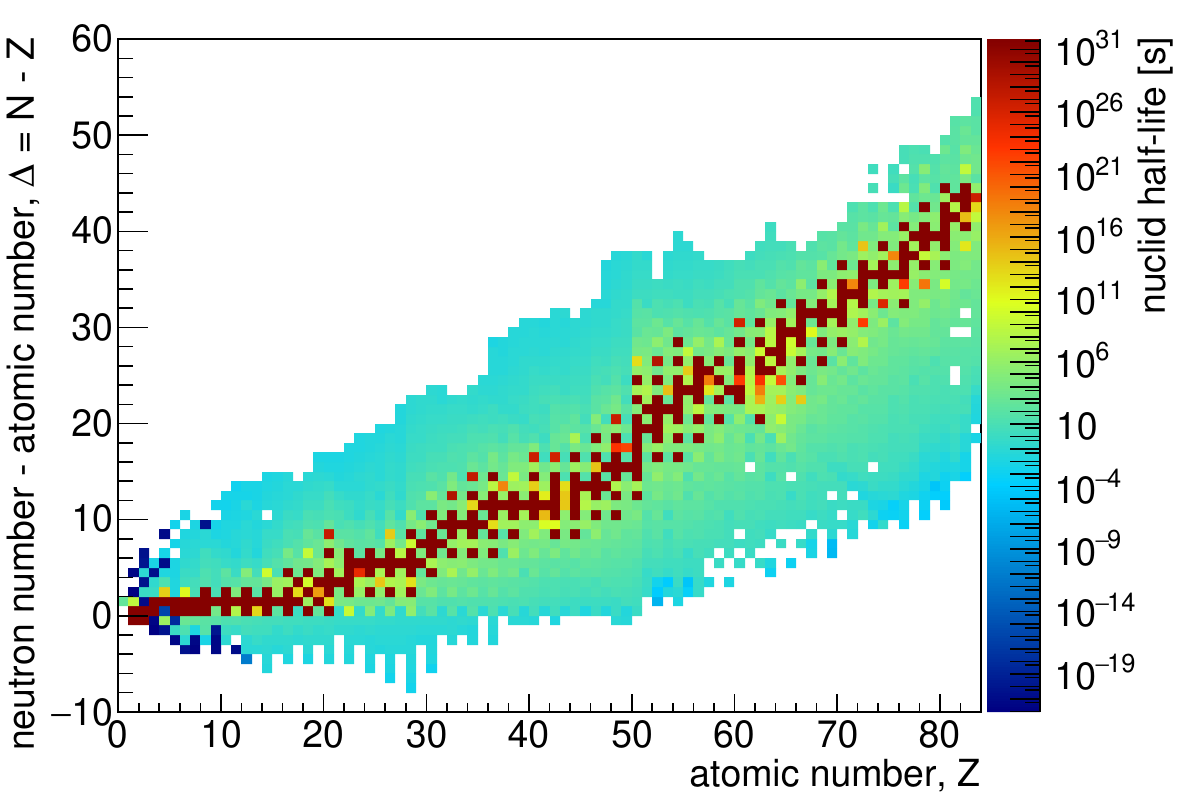}
  \caption{Half-life of known nuclei \cite{nubase}.}
  \label{fig:decay}
\end{figure}

Since the detectors are positioned quite far away from the interaction region, it is worth checking which produced nuclei have a chance to reach the AFP stations before they decay.
The spectators move with velocity close to the speed of light and their Lorentz factor is $\gamma=2751$  (for the lead beam and the LHC magnets set as for 6500 GeV proton beams).
Then, the proper time of a nucleus before it reaches the AFP detector is about 0.3~ns.
Fig. \ref{fig:decay} presents the half-life times of the known nuclei%
\footnote{This and other plots in this paper are presented in a non-standard way, as a function of the atomic number $Z$ and the difference of the number of neutrons, N, and the atomic 
number, $\Delta = N - Z = A - 2Z$, where $A$ is the mass number.}.
It is clear that the vast majority of the nuclei could reach the detectors before decaying.

Neglecting the internal motion of nucleons within the nucleus, all the nucleons carry the same energy, equal to the energy of the beam divided by the mass number of the beam particles: $E_N = E_b/A_b$.
Assuming that the spectators are left intact in the collision, a spectator fragment with mass number $A$ would have the energy equal to $A\cdot E_N$.

In order to take into account the internal motion of nucleons, the Fermi-gas model of a nucleus was employed.
In the rest frame of the nucleus, the density of nucleon states is given by $\text{d} n \sim p^2 \text{d} p$.
Therefore, in the simulation the absolute value of the momentum of each nucleon was randomly drawn from a quadratic distribution between zero and the Fermi momentum of 250~MeV.
The direction of momentum of each nucleon was assumed to be isotropically distributed.
The momentum of a given fragment was calculated as a vector sum of momenta of all its nucleons.
Finally, the nucleus momentum was Lorentz-transformed into the laboratory frame.

The transport of nuclear fragments was simulated using the Mad-X program.
The setting for the lead beams were deployed, obtained assuming that the beam of fully ionised \ce{^{208}_{82}$Pb$} ions is accelerated to the energy of 2.56 TeV per nucleon, which corresponds to 6.5 TeV energy protons.
The optics of $\beta^\ast = 0.8$~m~\cite{bib:lhcoptics} was taken, which corresponds to the LHC Run 2 heavy ion operations.
In order to simulate trajectories of ions different from lead, their momenta were scaled to the momentum of lead ion that would have the same trajectory in the magnetic field.
This procedure is possible because the trajectory in the magnetic field depends effectively on the ratio of the particle momentum to its charge.

Neglecting the spreads due to the beam emittance and the Fermi motion, a nucleus with a given $A$ and $Z$ numbers will hit the AFP detectors in a 
well defined position.
Since the dipole magnets bend the beam in the horizontal direction, then the $x$-coordinate of the nuclear fragment trajectory plays the major role in the present considerations.
One should recall that for the safety reasons the detector is positioned at some distance with respect to the beam and that there is additional dead material of the roman pot floor.

\begin{figure}[htbp]
  \centering
  \begin{minipage}{0.5\textwidth}(a)\end{minipage}%
  \begin{minipage}{0.5\textwidth}(b)\end{minipage}
  \includegraphics[page=4, width=0.49\linewidth]{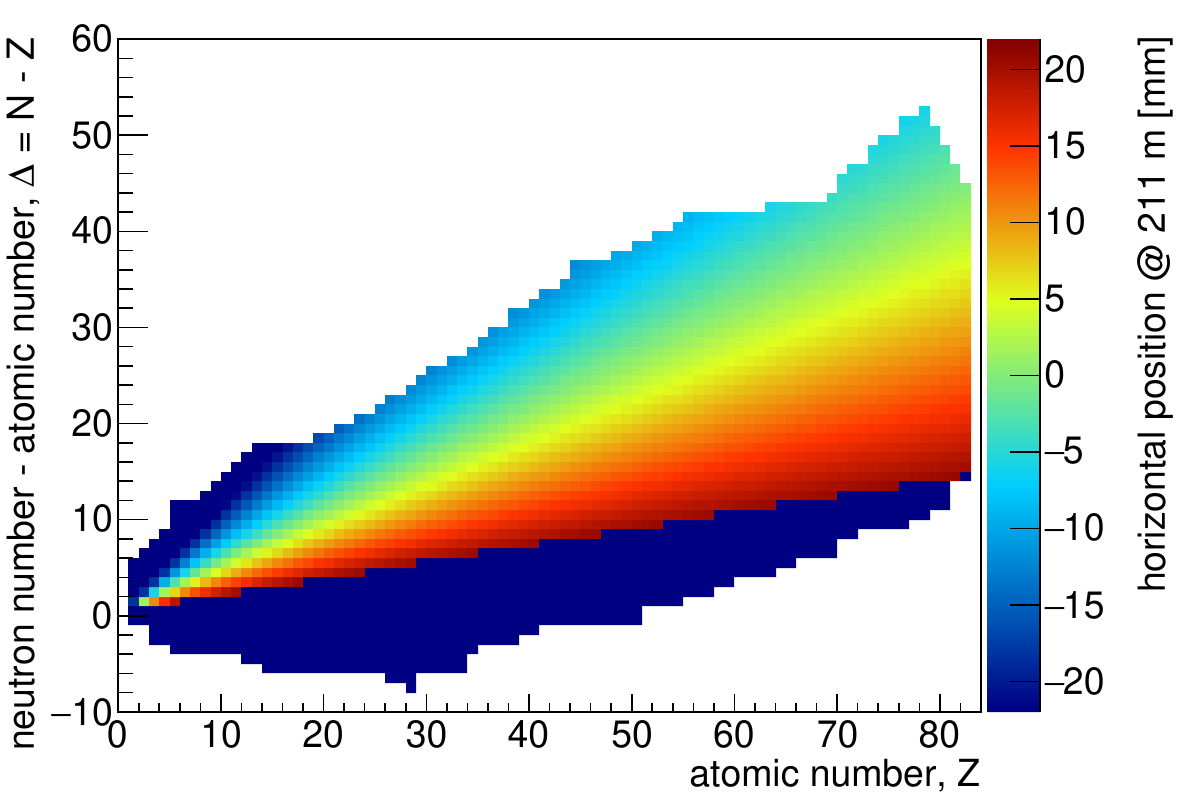} \hfill
  \includegraphics[page=5, width=0.49\linewidth]{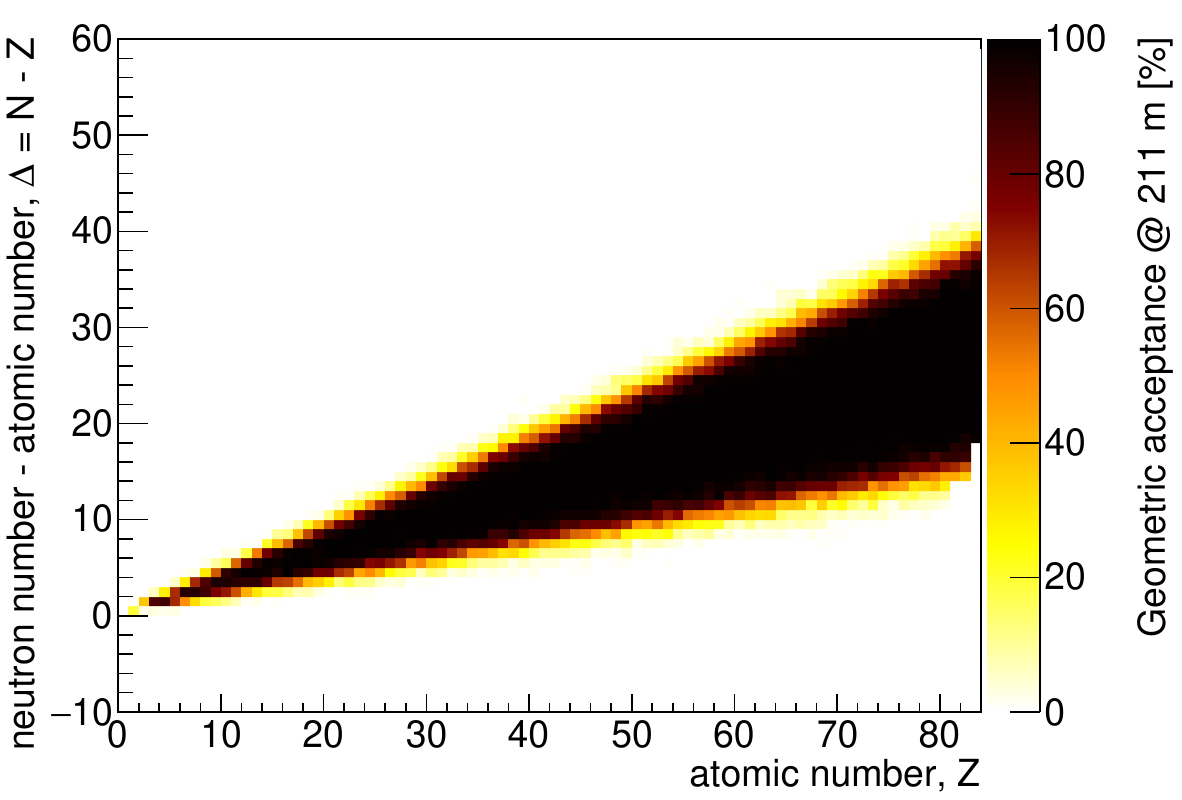}
  \caption{(a) Horizontal position of nuclei at 211 m from the ATLAS interaction point.
  The nuclei lost in the LHC apertures are drawn in dark blue. 
  (b)
  AFP acceptance for nuclear debris.}
  \label{fig:pos_acc}
\end{figure}

Fig. \ref{fig:pos_acc}a shows the horizontal positions of all nuclei at 211 m from the interaction point (in the middle between the two AFP stations).
Naturally, the position of \ce{^{208}_{82}$Pb$} ($\Delta=44$) and all nuclei with the same $A/Z$ ratio (\textit{i.e.} the same $\Delta/Z$ ratio) is equal to zero.
Nuclei with less neutrons per proton  are deflected outside the LHC ring, similarly to diffractive protons, and can be registered in the AFP detectors.
Nuclei containing more neutrons per proton are deflected towards the LHC centre escaping the detection.
Nuclei with $A/Z$ very different from that for lead can be lost in the LHC apertures and not reach 211 m at all.

For the emittance value of 1.233~$\upmu$m \cite{bib:lhcoptics}, the lead beams at the interaction point have the angular spread of 24~$\upmu$rad,
while the interaction vertex distribution has the transverse spread of 13~$\upmu$m
and the longitudinal one of 5.5~cm.
The horizontal width of the beam at 211~m from the interaction point, $\sigma_x$, is 134~$\upmu$m.
This width is a usual unit of the distance between the detector and the beam. 
In the following, it was assumed that the sensitive area of the sensor is placed 3~mm from the centre of the beam.
This distance covers about $19\sigma_x$ and a 0.5 mm-long distance between the active detector edge and outer wall of the roman pot floor.

Fig. \ref{fig:pos_acc}b 
shows the acceptance to detect a given nucleus as a function of $Z$ and $\Delta$.
The obtained results were averaged over the distributions of momenta discussed before and the Gaussian spreads of the LHC beams (spatial and angular).

Although, the AFP detectors were not thought to detect the nuclear debris,  their acceptance covers a significant part of the nuclei spectrum.
This is particularly true for the heavier nuclei, where for a given $Z$ more than a half of known nuclei can be potentially detected.
With decreasing $Z$ value the range of the accepted masses  linearly decreases.

\begin{figure}[h]
  \centering
  \begin{minipage}{0.5\textwidth}(a)\end{minipage}%
  \begin{minipage}{0.5\textwidth}(b)\end{minipage}
  \includegraphics[width=0.5\linewidth]{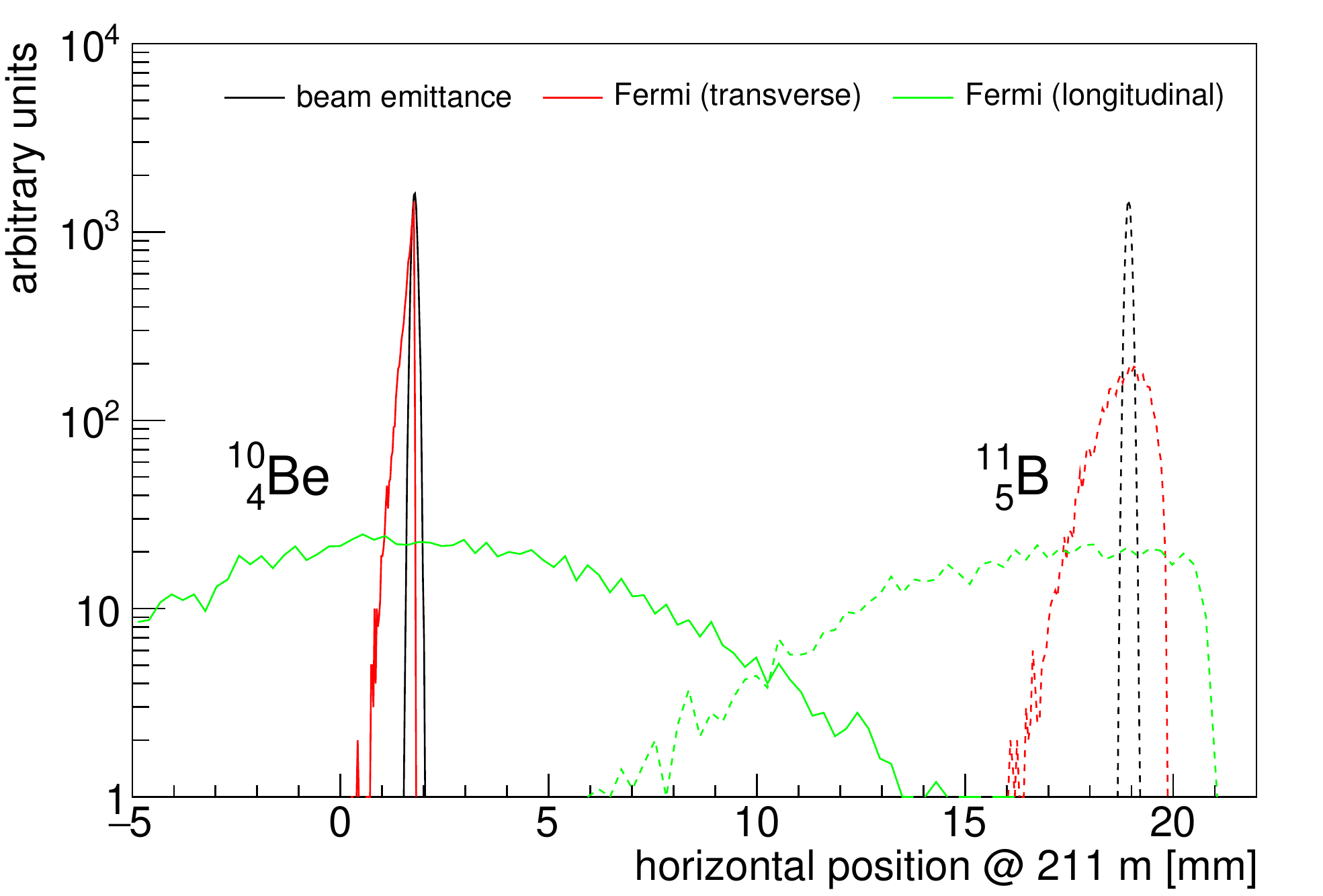}%
  \includegraphics[width=0.5\linewidth]{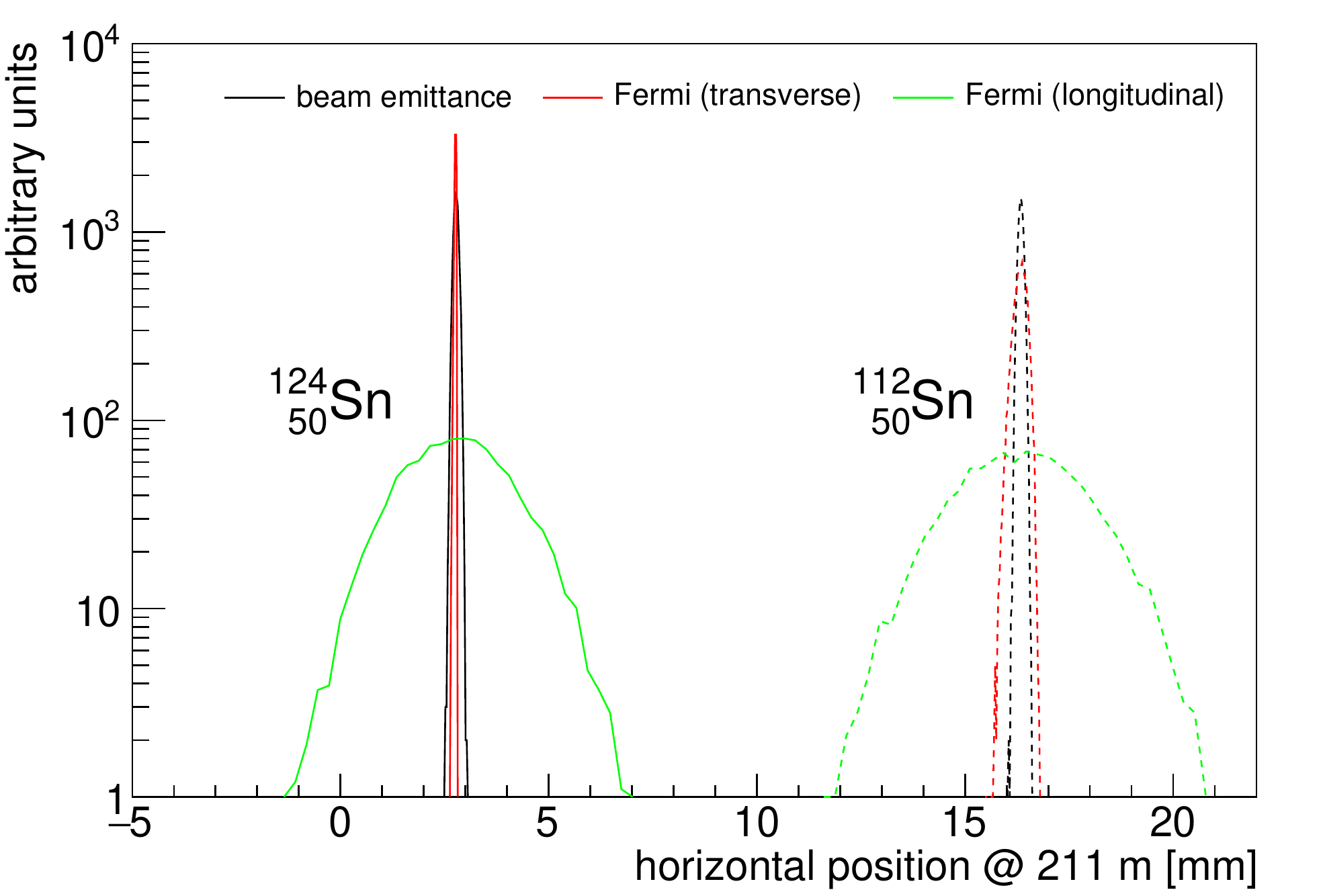}
  \caption{Effects of finite beam emittance and the Fermi motion on the fragment position in AFP for exemplary light (a) and heavy (b) nuclei.}
  \label{fig:effects}
\end{figure}

A study of the influence of various spreads on the position of selected ions at the distance of 211 m away from the interaction point is presented in Fig.~\ref{fig:effects}.
As can be observed, the effects of the beam spreads and those due to the transverse component of the Fermi motion are quite small.
The position smearing is dominated by the longitudinal Fermi motion magnified by the Lorentz boost. One can observe that this effect is stronger for lighter nuclear debris. 

\section{Towards the centrality determination}
\label{sec:dpmjet}

\begin{figure}[b]
  \centering
  \includegraphics[width=0.5\textwidth, page=1]{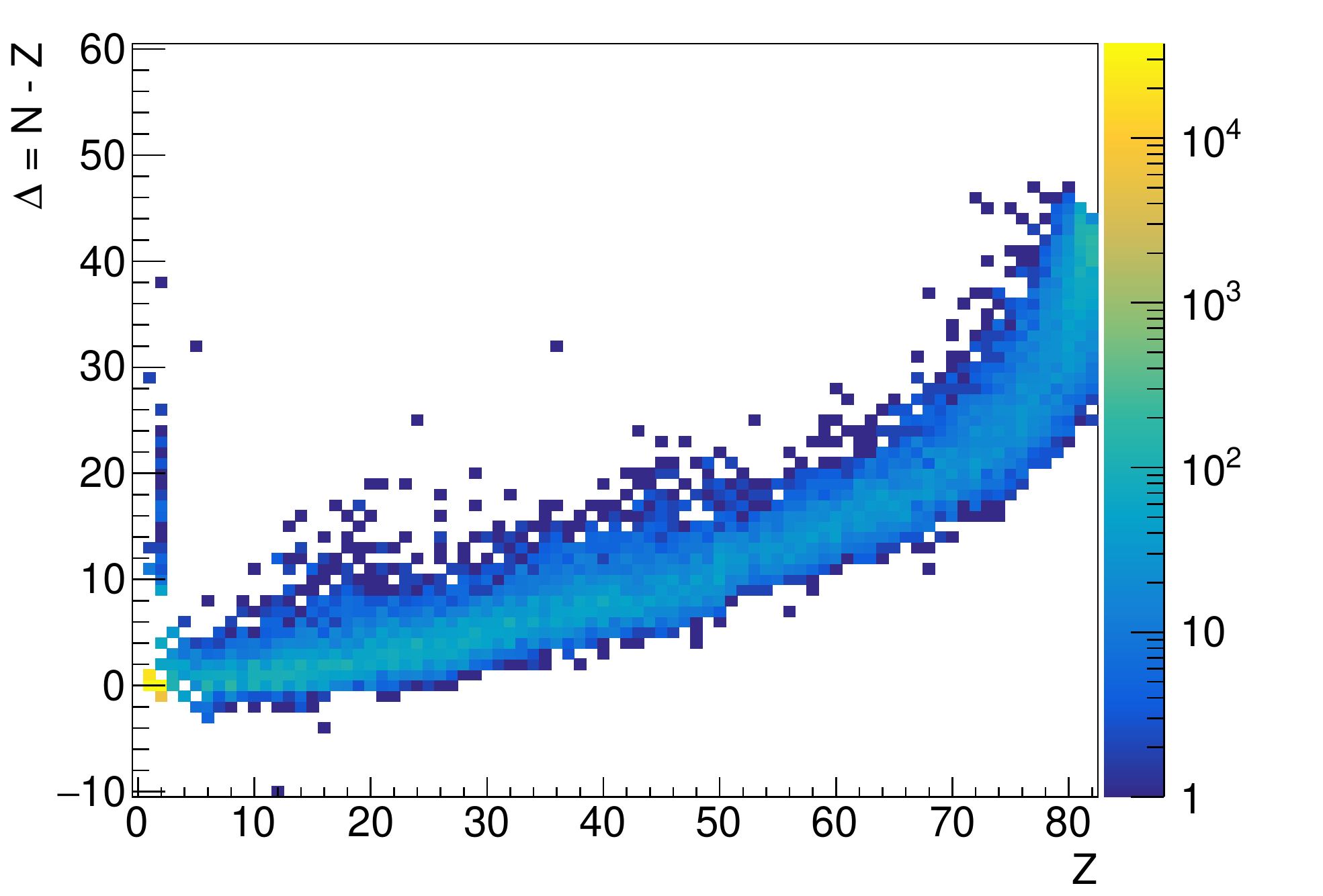}
  \caption{Abundance of nuclear fragments produced in PbPb collisions simulated using DPMJET.}
  \label{fig:abundance}
\end{figure}

The discussion concerning the details of the fragmentation process is outside the scope of the present study, see for example \cite{ferrari, dpmjet, olson, frankland,jacak} and references therein.
In order to study how the measurement of fragments can be used to get information about the central state, a simulation of PbPb collisions using the DPMJET Monte Carlo event generator \cite{Roesler:2000he} was performed.
For each simulated event, the generator reports a list of produced particles, including the spectator fragments.
The distribution of the produced fragments is presented in Fig.~\ref{fig:abundance}.

\begin{figure}[htbp]
  \centering
  \begin{minipage}{0.5\textwidth}(a)\end{minipage}%
  \begin{minipage}{0.5\textwidth}(b)\end{minipage}
  \includegraphics[width=.5\textwidth, page=4]{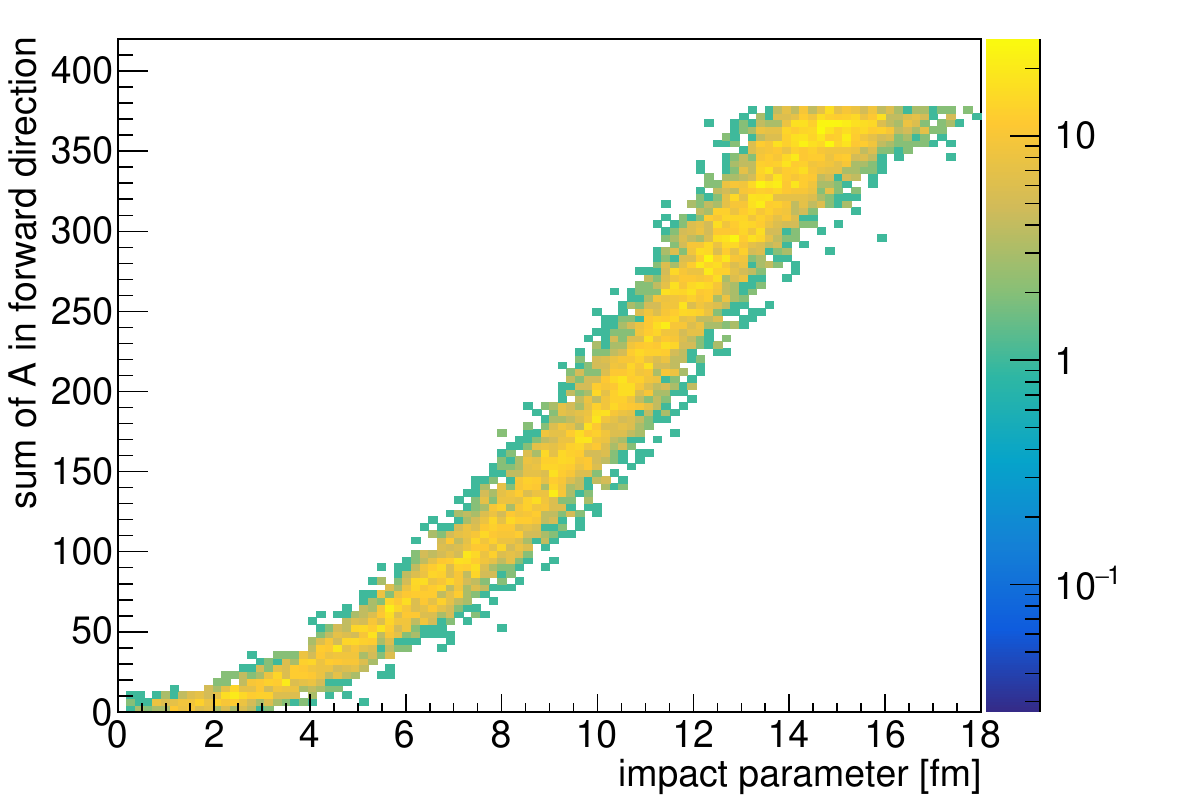}\hfill
  \includegraphics[width=.5\textwidth, page=7]{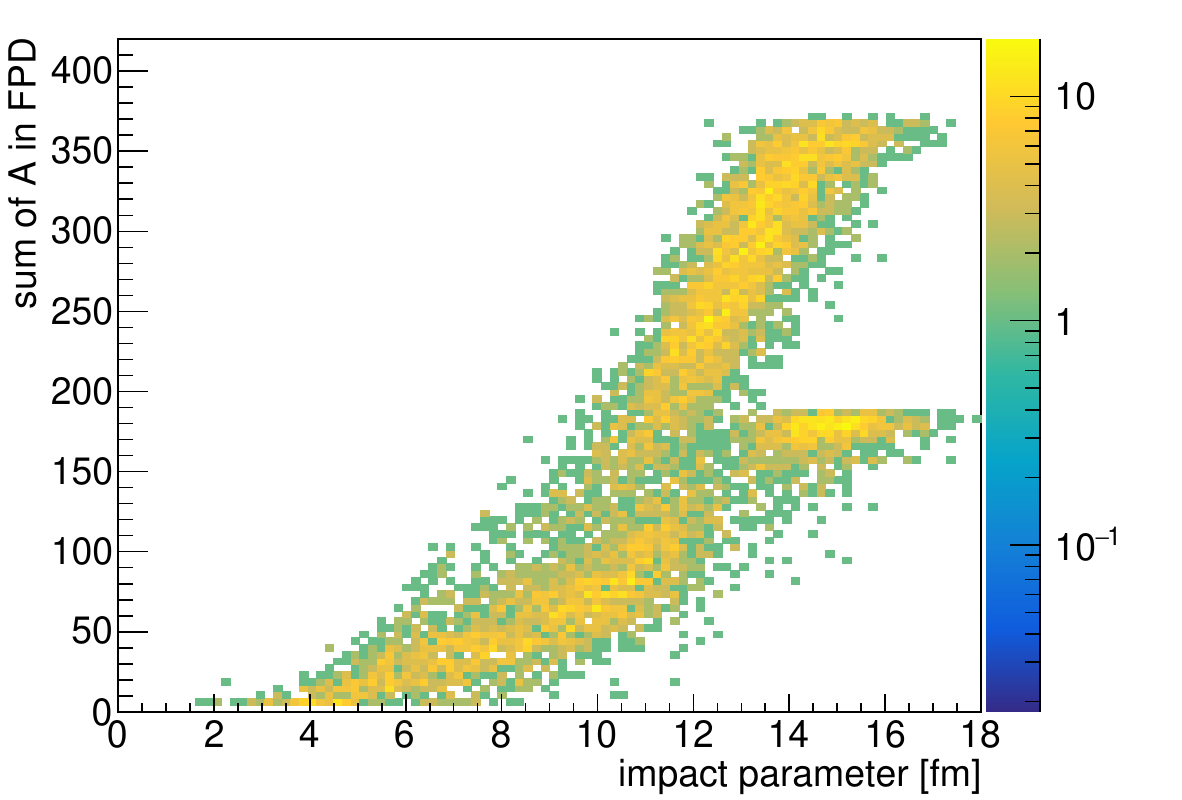}
  \caption{Correlation between the impact parameter of the collision and the sum of mass numbers of: (a) all produced nuclear fragments, (b) nuclear fragments within the acceptance of forward proton detectors.}
  \label{fig:impact}
\end{figure}

Each event is generated with a known, random value of the impact parameter.
Fig.~\ref{fig:impact}a presents the correlation between its value and the sum of mass numbers of all\footnote{For technical reasons, free protons and neutrons are not counted.} produced nuclear fragments.
As expected, there is a strong dependence between these two variables -- the more peripheral the event, the more spectators are produced.
However, not all fragments are within the acceptance of the detectors.

Fig.~\ref{fig:impact}b shows the same correlation, but here the sum goes only over the fragments registered in the forward proton detectors (FPD)%
\footnote{For the results based on the DPMJET simulation, the acceptance of forward proton detectors is taken into account in an approximate way based on their $A/Z$ ratio.}.
One can see that some correlation is still present, but it is not as strong as before.
In addition, it looks like consisting of two different correlations added together -- one correlation similar to the original one, while the other one with $\sum A$ scaled down.

\begin{figure}[htbp]
  \begin{minipage}{0.5\textwidth}(a)\end{minipage}%
  \begin{minipage}{0.5\textwidth}(b)\end{minipage}
  \includegraphics[width=.5\textwidth, page=9]{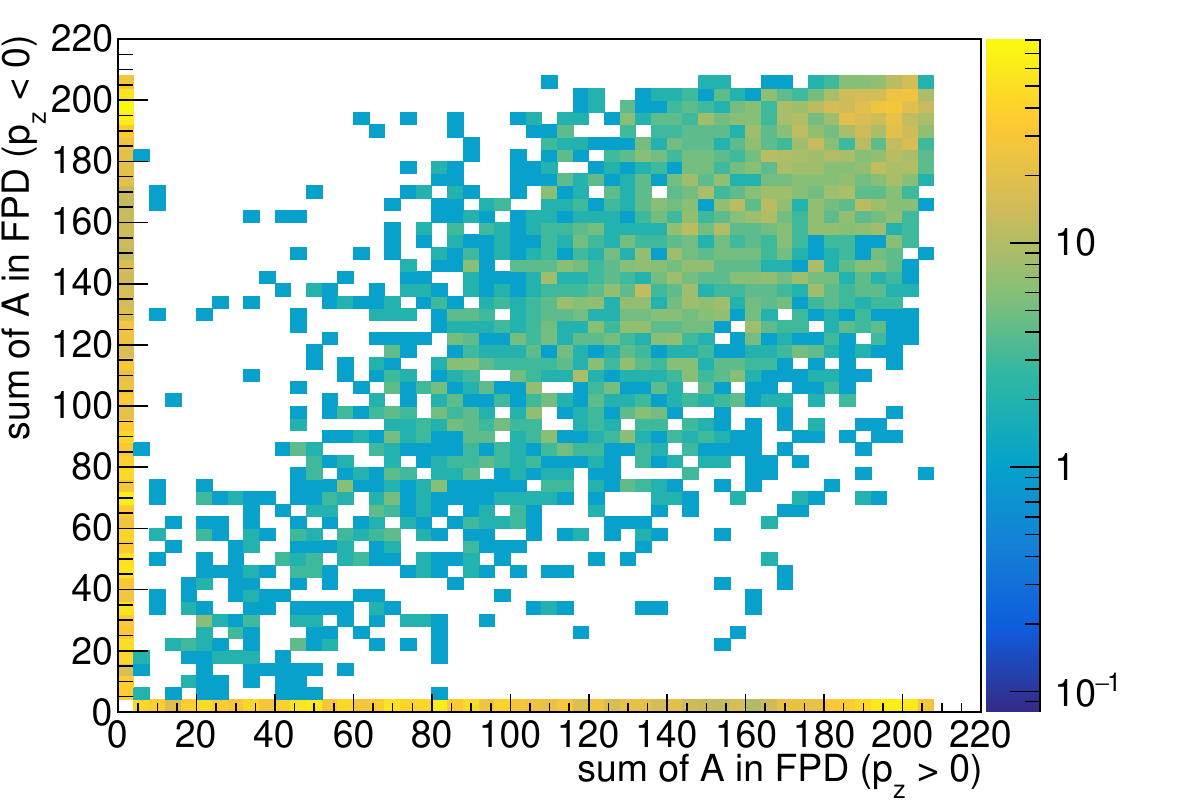}\hfill
  \includegraphics[width=.5\textwidth, page=8]{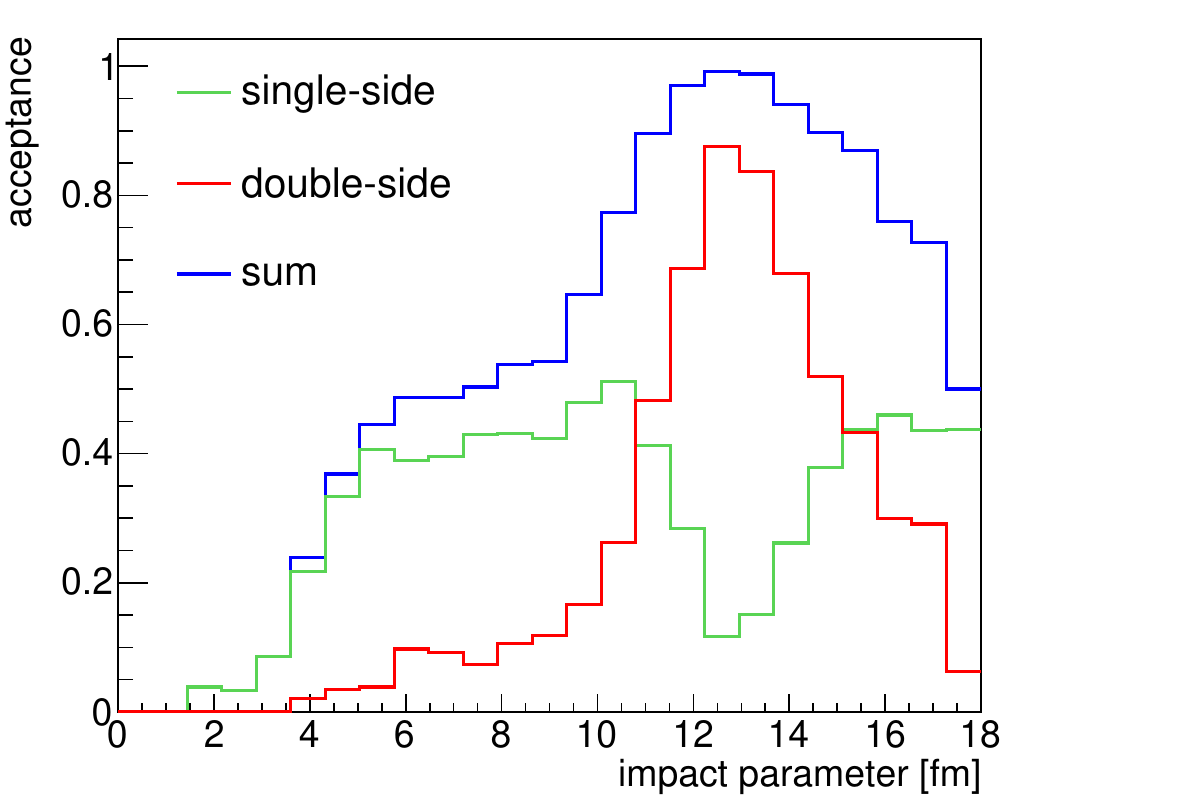}
  \caption{
  (a)
  Correlation between the sum of mass numbers of nuclear fragments within the acceptance of forward proton detectors on the sides with positive and negative longitudinal momentum. 
  (b)
  Acceptance for events with nuclear fragments being measured in forward proton detectors.}
  \label{fig:asymmetry}
\end{figure}

The initial state of the PbPb collision is to a first approximation symmetric with respect to the $p_z$ sign.
However, the fluctuations of the ions shape can break this symmetry.
In addition, fragmentation of each of the spectator systems is a random process and their fluctuations are independent (at least there is no obvious reason for the existence of a correlation).
All this can cause an asymmetry between the fragments produced with positive and negative $p_z$ values.
This is illustrated in Fig.~\ref{fig:asymmetry}a showing the correlation between the sum of $A$ measured on the two sides.
Based on this result, two classes of events can be distinguished: events with nuclear fragments measured on both sides (double tag) and only on one side (single tag). 
It is worth mentioning that for the double-tag events, the $\sum A$ of the fragments measured on both sides are correlated.
The width of this correlation reflects the correlation between the impact parameter (common for the whole event) and the measurements on each side.

It is interesting to ask what is the probability that a given event will be of either type.
This value depends on the impact parameter and is depicted in Fig.~\ref{fig:asymmetry}b.
For the most central events, the probability of observing any fragments in the forward detectors is to a good approximation zero.
This comes from the fact that in such collisions only the lightest spectator fragments, which will escape the registration, can be formed.
With increasing impact parameter, the probability of single-tag events increases.
But at about 12 -- 14 fm, it drops down, which corresponds to a peak in the probability for double-tag events.

\begin{figure}[htbp]
  \begin{minipage}{0.5\textwidth}(a)\end{minipage}%
  \begin{minipage}{0.5\textwidth}(b)\end{minipage}
  \includegraphics[width=.5\textwidth, page=6]{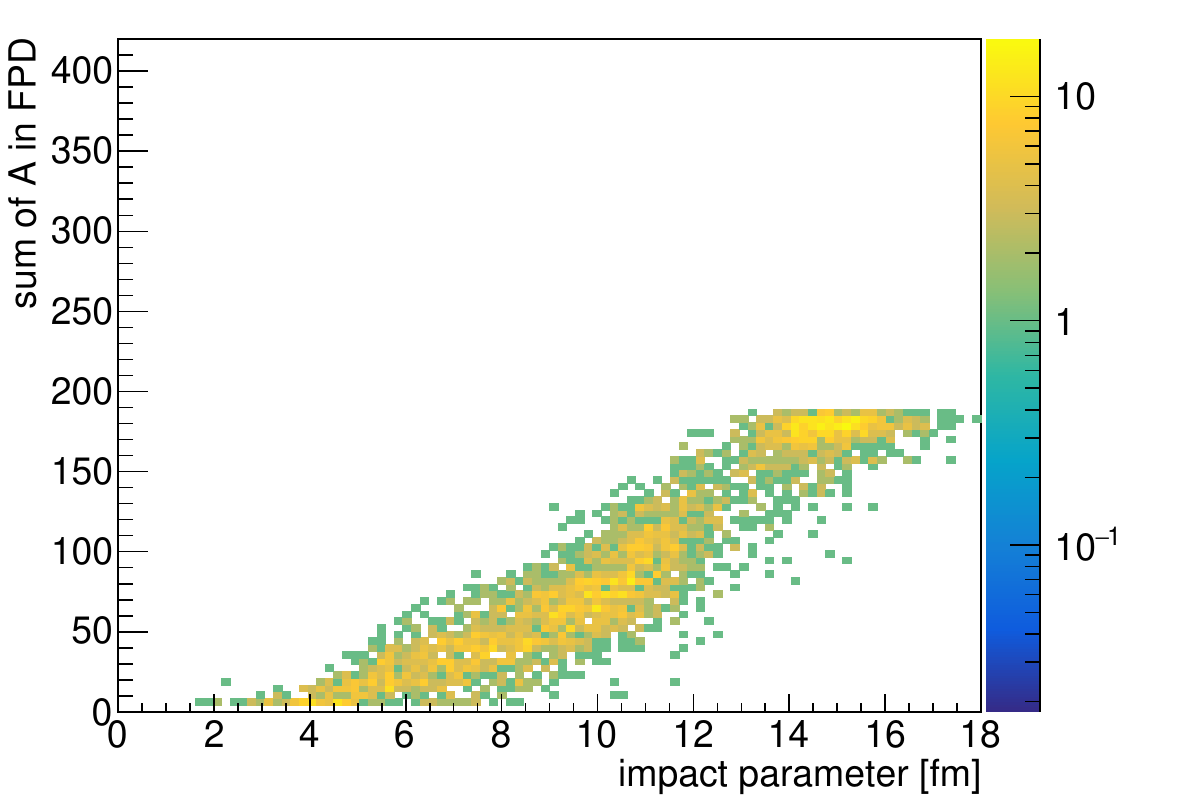}\hfill
  \includegraphics[width=.5\textwidth, page=5]{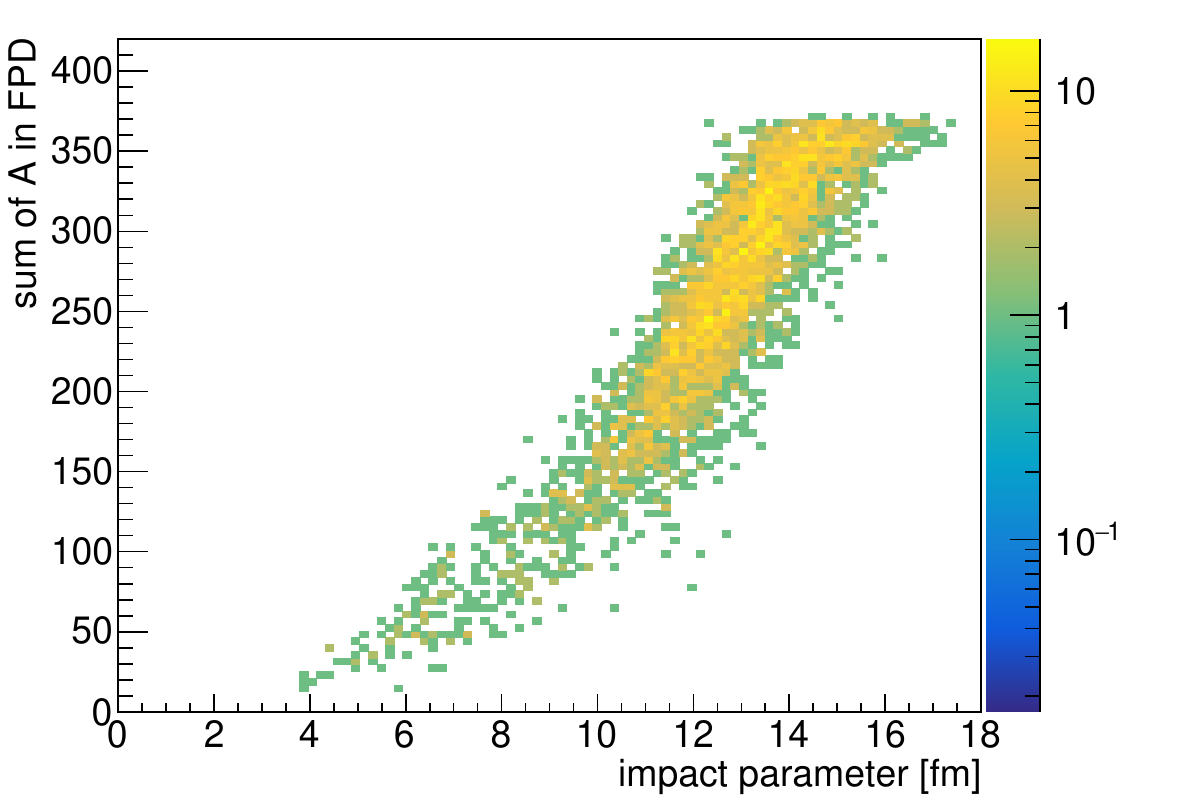}
  \caption{Correlation between the impact parameter of the collision and the sum of mass numbers of nuclear fragments within the acceptance of forward proton detectors for (a) single-tag and (b) double-tag events. }
  \label{fig:corr}
\end{figure}

Fig.~\ref{fig:corr} presents the correlations between the impact parameter and the sum of $A$ for measured fragments for the two types of events separately.
A~correlation between the two variables is visible in both cases, which demonstrates that the proposed method can be used for centrality determination.

\section{Summary and conclusions}
\label{sec:conclusions}

In the presented study it was shown that the existing forward proton detectors at the LHC provide an interesting possibility of detecting 
nuclear debris emerging from the collision of two heavy ions.
This possibility could be used for a measurement of abundances of various fragments produced in heavy-ion collisions.

Another interesting goal could be a measurement of the centrality of PbPb collisions.
Different centralities result in different signals generated by the produced nuclear debris.
Such a measurement would be independent of and complementary to other commonly used methods, which are based on measurements of particles originating from the interactions of participant nucleons.
With several forward detectors that would provide large acceptance, a direct measurement of 
the number of spectators and hence the determination of the number of participants could be possible, see also~\cite{Tarafdar:2014oua}.
However, the present work shows that one can get information about the centrality even with the limited acceptance of already existing detectors.

Finally, it is worth mentioning that measurements of spectators using roman pot detectors could be considered not only for ion--ion collisions, but also for proton--ion or lepton--ion interactions.
However, the fragmentation of a nucleus excited by a proton or a virtual photon can be very different from the fragmentation of the spectator system discussed in this paper.
Therefore, this topic requires a dedicated study to understand what could be a possible physics motivation for such measurements and what experimental setup would be needed.

\section*{Acknowledgements}
We gratefully acknowledge discussions with S. Tarafdar and A. Milov on these topics.
This work was supported in part by Polish National Science Center grant no. 2015/19/B/ST2/00989.

\end{document}